\documentclass[11pt,a4paper,english,3p,number,sort&compress]{elsarticle}
\usepackage{ae,aecompl}
\usepackage[T1]{fontenc}
\usepackage[latin9]{inputenc}
\usepackage{color}
\usepackage{xcolor}
\usepackage{amsmath}
\usepackage{amssymb}
\usepackage{graphicx}
\usepackage{float}
\newtheorem{thm}{Theorem}
\newtheorem{lem}[thm]{Lemma}
\newdefinition{rmk}{Remark}

\newenvironment{pf}{\textbf{Proof:}}{\linespread{1.66}\hspace{\stretch{1}}{$\square$}}

\newproof{pot}{Proof of Theorem \ref{thm2}}

\makeatletter

\newcommand{\ds}{\displaystyle}

\makeatother

\usepackage{babel}
\begin{document}
\begin{frontmatter}
\title{A generalised distributed-order Maxwell model}

\author[UM]{L.L. Ferr\'as}
\ead{luislimafr@gmail.com}

\author[LM]{M.L. Morgado}
\ead{luisam@utad.pt}

\author[MR]{M. Rebelo\corref{cor1}} 
\ead{msjr@fct.unl.pt}


\cortext[cor1]{Corresponding author}

\address[UM]{Department of Mechanical Engineering - FEUP, University of Porto \& CMAT - Centre of Mathematics, University of Minho, Portugal.}

\address[LM]{CEMAT, Instituto Superior T{\'e}cnico, Universidade de Lisboa \& Department of Mathematics, University of Tr\'as-os-Montes e Alto Douro, UTAD, Vila Real, Portugal.}

\address[MR]{Centro de Matem{\'a}tica e Aplica\c c\~oes  (CMA) and Departamento de 
Matem{\'a}tica, Faculdade de Ci\^encias e
Tecnologia, Universidade NOVA de Lisboa,  Quinta da Torre, 2829-516 Caparica, 
Portugal.}

\begin{abstract} 
In this work we present a generalised viscoelastic model using distributed-order derivatives. The model consists of two distributed-order elements (distributed springpots) connected in series, as in the Maxwell model. The new model generalises the fractional viscoelastic model presented in [H. Schiessel, A. Blumen, Hierarchical analogues to fractional relaxation equations, Journal of Physics A: Mathematical and General 26 (1993) 5057-50] and allows for a more broad and accurate description of complex fluids when a proper weighting function of the order of the derivatives is chosen.
We discuss the connection between classical, fractional, and viscoelastic models of distributed order and highlight the fundamental concepts that support these constitutive equations.
We also derive the relaxation modulus, the storage and loss modulus, and the creep compliance for specific weighting functions. 

\end{abstract}  
\end{frontmatter}

\section{Introduction}
\noindent

Viscoelastic fluids play an important role in our daily lives. Therefore, it is essential to derive accurate models that can describe and predict the behaviour of such complex fluids under different types of deformations.

The first widely used viscoelastic models date back to 1876 with the works of Boltzmann \cite{Boltzmann2,Boltzmann3} and Maxwell \cite{Markovitz,Kohlrausch,Maxwell1867},

\begin{equation}
\sigma\left(t\right)=\intop_{0}^{t}G_{0}e^{-\frac{t-t'}{\lambda}}\frac{d\gamma}{dt'}dt'\label{eqmaxintegral}
\end{equation}

where $\sigma(t)$ is the stress, $G\left(t\right)=G_{0}e^{\frac{-t}{\lambda}}$ is the relaxation modulus, $\gamma(t)$ is the deformation or strain, $G_0$ and $\lambda$ are two fitting parameters to be obtained from experimental results.

By differentiating both sides of (\ref{eqmaxintegral}) with respect to time, we easily obtain the following differential form of the model:

\begin{equation}
\sigma(t)+\frac{\eta}{G_{0}}\frac{d\sigma(t)}{dt}=\eta\frac{d\gamma(t)}{dt},\label{maxdiff}
\end{equation}
with $\eta=\lambda G_{0}$.

\begin{figure}[!ht]
\centering
\begin{tabular}{cc}  \includegraphics[scale=0.24]{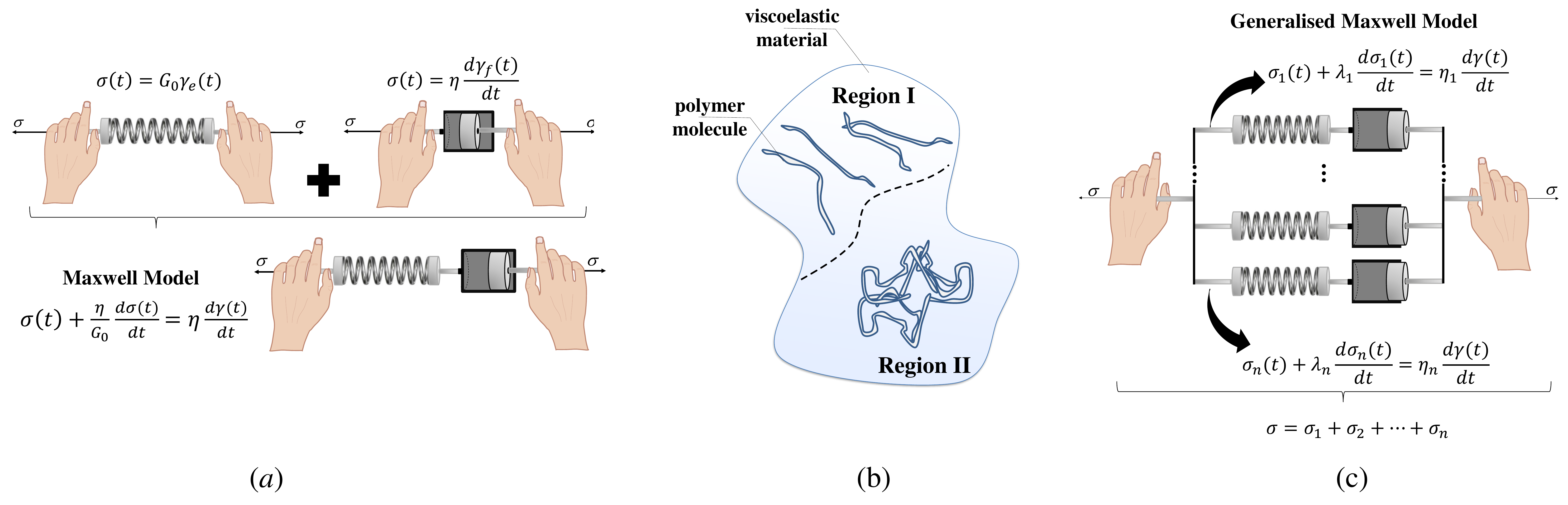}
\end{tabular}
\caption{(a) Mechanical analogue of the Maxwell model. Assuming the total rate of deformation is given by the sum of the viscous and elastic contributions, then the Maxwell model is obtained; (b) Different states of relaxation in the same material; (c) The generalised Maxwell Model.}
\label{maxmech}
\end{figure}

In 1903 J.H. Poynting and  J.J. Thomson \cite{Poynting} introduced  the \emph{spring and damper (or dashpot)} analogy. They introduced the Maxwell model by using a spring and a dashpot, as shown in figure \ref{maxmech}(a) (for a more detailed explanation see \cite{ferrasbook, FerrasNT2017, ferras2018, ferras2019} and for the derivation of the Maxwell model using molecular theory see \cite{Bird1987,Huilgol1997,Larson}).

Figure \ref{maxmech}(b) shows a case where the Maxwell model may fail to correctly predict the behaviour of a viscoelastic fluid. When the same material shows different relaxations in different regions, the model is not able to deal with such complexity. Some solutions were proposed in  the literature to deal with this problem, such as the use of Prony series: the different regions of the material are modelled by using more than one Maxwell model (or other viscoelastic model), each one adding up a contribution to the total stress observed in that material (see figure \ref{maxmech}(c)).

\subsection{Single-Order Fractional Viscoelastic Models}

Instead of using a Prony series to obtain a better model, one can also use different relaxation functions. For example,  the Maxwell-Debye relaxation (exponential decay) is observed in several complex viscoelastic fluids, but there are other materials showing different types of fading memory, such as an algebraic decay, $G\left(t\right)=St^{-\alpha}$, with $0<\alpha<1$ and $S$ a scalar measure of the strength of the material \cite{Bavand2016}.

If we re-write the relaxation modulus in the form $G\left(t-t'\right)=\frac{\mathbb{V}}{\Gamma\left(1-\alpha\right)}\left(t-t'\right)^{-\alpha}$, then, equation (\ref{eqmaxintegral}) can be written as, 
\begin{equation}
\boldsymbol{\sigma}\left(t\right)=\frac{1}{\Gamma\left(1-\alpha\right)}\intop_{0}^{t}\mathbb{V}\left(t-t'\right)^{-\alpha}\frac{d\boldsymbol{\gamma}}{dt'}dt'.
\end{equation}
where $\mathbb{V}$ is a constant for a fixed $\alpha$, with physical dimensions $Pa.s^{\alpha}$. $\mathbb{V}$ is a generalised modulus or a \emph{quasi-property} \cite{Dingle} (see also \cite{Dingle,Jaishankar2014,Scott1947,Koeller1984,Schiessel1993,FerrasNT2017,Schiessel1995,Friedrich}).

\begin{figure}[!ht]
\centering
\begin{tabular}{cc}  \includegraphics[scale=0.18]{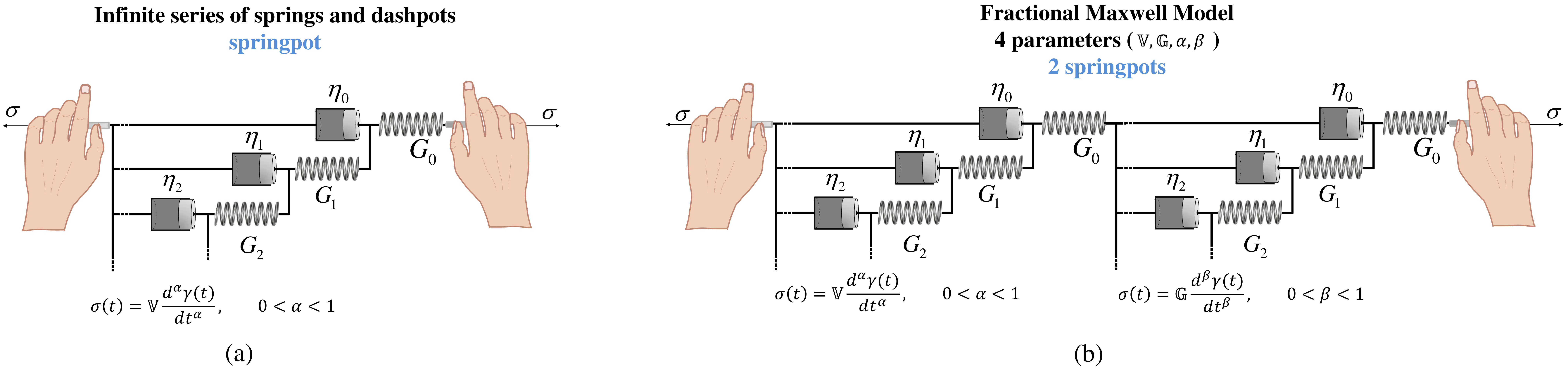}
\end{tabular}
\caption{The springpot (a) and the Fractional Maxwell Model (b).}
\label{FMM}
\end{figure}

The fractional derivative in the Caputo sense ($0<\alpha<1$) is given by \cite{Caputo1969}:
\begin{equation}
_{0}^{C}D_{t}^{\alpha}f\left(t\right)=\frac{1}{\Gamma\left(1-\alpha\right)}\intop_{0}^{t}\left(t-t'\right)^{-\alpha}\frac{df}{dt'}dt'.\label{eqcaputo}
\end{equation}
The constitutive equation for a material exhibiting relaxation $St^{-\alpha}$ can therefore be re-written as $\boldsymbol{\sigma}\left(t\right)=\mathbb{V}~_{0}^{C}D_{t}^{\alpha}\boldsymbol{\gamma}\left(t\right)$. Using the compact notation $_{0}^{C}D_{t}^{\alpha}\equiv\frac{d^{\alpha}}{dt^{\alpha}}$, we have, $\boldsymbol{\sigma}\left(t\right)=\mathbb{V}\frac{d^{\alpha}\boldsymbol{\gamma}\left(t\right)}{dt^{\alpha}}$. This model is illustrated in figure (\ref{FMM}) by a springpot (a combination of springs and dashpots in a ladder structure - see \cite{Schiessel1993,Schiessel1995} for the derivation of the mechanical analogue, the physical meaning of $\eta_i$ and $G_i$, $i=0,1,2,...$, and, the restrictions imposed on these parameters).

Two springpots arranged in series lead to the Fractional Maxwell Model \cite{Schiessel1995,Friedrich},
\begin{equation}
\boldsymbol{\sigma}\left(t\right)+\frac{\mathbb{V}}{\mathbb{G}}\frac{d^{\alpha-\beta}\boldsymbol{\sigma}\left(t\right)}{dt^{\alpha-\beta}}=\mathbb{V}\frac{d^{\alpha}\boldsymbol{\gamma}\left(t\right)}{dt^{\alpha}},\label{eqfracmaxwell}
\end{equation}
\noindent where it has been assumed (without loss of generality) that $0<\beta\leq\alpha< 1$. $\mathbb{G}$ is also a \emph{quasi-property}. This four parameter model is able to describe a much wider range of complex fluid behaviour when compared to the classical Maxwell model (obtained in the limit $\alpha =1,\beta =0$).

\subsection{Distributed-Order Fractional Viscoelastic Models}

More recently, inspired by the works of Bagley, Torvik, Atanackovic, and Caputo \cite{Atanackovic2002, Atanackovic2003, Bagley2000a,Bagley2000b,Caputo2001}, the research group has developed a model that generalises the classical and fractional viscoelastic models. \\
Following \cite{symcomp2021} the stress can be written as \begin{eqnarray}\label{sigma}
\sigma\left(t\right)=\gamma_0G(t)H(t)+\intop_{0^+}^{t}G\left(t-t'\right)\frac{d\gamma}{dt'}dt', 
\end{eqnarray}
where $\gamma$ is the strain and $H(t)$ is the Heaviside function. All events over the history of a viscoelastic material contribute to the current state of stress and strain, and therefore we should consider the lower limit to be $-\infty$. In this work, for ease of understanding, we assume that the state of stress at $t=0$ is not influenced by any kind of past ($t<0$) process (polymerization, heating, etc) that would create changes to the molecular structure and therefore influence the present and future (for $t\geq 0$).
\\
Recall that there are materials showing different types of fading memory, such as an algebraic decay, $G\left(t\right)=St^{-\alpha}$. Now assume that a material shows a combination of algebraic decays, such as,
\begin{equation}
G\left(t\right)=S_1t^{-\alpha_1}+S_2t^{-\alpha_2}+...+S_nt^{-\alpha_n},
\end{equation}
with $0<\alpha_i<1$, $i=1,...,n$.
\\ Assuming that $n\rightarrow \infty$, covering the open set $(0,1)$, then we can write a continuous version of the previous finite combination of algebraic decays, which is given by:
\begin{equation}
G\left(t\right)=\int_0^1 S(\alpha)t^{-\alpha}~d\alpha,
\end{equation}
where $S(\alpha)$ is a function that interpolates $(\alpha_i,S_i)\, i=1,...,n.$.\\
Therefore, let $c(\alpha)$ be a function such that  $S(\alpha)=c(\alpha)/\Gamma(1-\alpha)$ ($c(\alpha)$ will be defined later), then, the relaxation modulus can  be written in the form:
\begin{equation}
G\left(t-t'\right)=\int_0^1 \frac{c(\alpha)}{\Gamma(1-\alpha)}(t-t')^{-\alpha}~d\alpha.\label{relaxd}
\end{equation}
 
Inserting equation (\ref{relaxd}) into (\ref{sigma}), for $t>0$ we obtain:
\begin{eqnarray}\nonumber 
\sigma\left(t\right)&=&
\gamma_0\int_0^1 \frac{c(\alpha)}{\Gamma(1-\alpha)}t^{-\alpha}~d\alpha+\int_0^t \int_0^1\frac{c(\alpha)}{\Gamma(1-\alpha)}(t-t')^{-\alpha}~d\alpha\,\gamma'(t')dt'
\\
\label{dist}&=&\int_0^1c(\alpha)\left( \frac{1}{\Gamma(1-\alpha)}\gamma_0 t^{-\alpha}~d\alpha+\frac{1}{\Gamma(1-\alpha)}\int_0^t (t-t')^{-\alpha}\gamma'(t')~dt'\right)d\alpha.
\end{eqnarray}
From \cite{DiethelmBook} we know that, $$\frac{1}{\Gamma(1-\alpha)}\gamma_0 t^{-\alpha}~d\alpha+\frac{1}{\Gamma(1-\alpha)}\int_0^t (t-t')^{-\alpha}\gamma'(t')~dt'=\ds_{0}^{RL}D_{t}^{\alpha}(\gamma(t)),$$ where $\ds _{0}^{RL}D_{t}^{\alpha}(\cdot)$ is the Riemann-Liouville differential operator:
\begin{eqnarray}
    \label{RLDef} \ds_{0}^{RL}D_{t}^{\alpha}(f(t))&=&\frac{d}{dt}\frac{1}{\Gamma(1-\alpha)}\int_0^t\frac{f(s)}{(t-s)^{\alpha}}ds\\
    \nonumber&=&\frac{f(0)t^{-\alpha}}{\Gamma(1-\alpha)}+\frac{1}{\Gamma(1-\alpha)}\int_0^t\frac{f'(s)}{(t-s)^{\alpha}}ds.
\end{eqnarray}

In a more compact form, from (\ref{dist}) we have that
\begin{equation}\label{DODEvisc}
\sigma\left(t\right)=_{0}^{RL}\mathbb{D}_{t}\left(\gamma(t)\right),
\end{equation}
where $\ds _{0}^{RL}\mathbb{D}_{t}$ is the Riemann-Liouville  distributed-order fractional derivative given by:
\begin{equation}
_{0}^{RL}\mathbb{D}_{t}f(t)=\int_0^1c(\alpha) \ds_{0}^{RL}D_{t}^{\alpha}f\left(t\right)~d 
\alpha\label{distorder}
\end{equation}
with $\ds_{0}^{RL}D_{t}^{\alpha}f\left(t\right)$  defined by (\ref{RLDef}). The function $c(\alpha)$, acting as a weight for the order of differentiation, is such that $ \displaystyle c(\alpha )\ge 0~~ \mbox{and}~~\int_0^1c(\alpha)~d \alpha=C>0$ (see \cite{Mainardib,Gorenflo2013} for more details).

\vspace*{0.1cm}

The function $c(\alpha)$ is used to represent mathematically the presence of multiple memory formalisms. 
If $\displaystyle{c(\alpha)=\delta(\alpha-\beta)}$, where $\delta (\cdot)$ is the delta Dirac function, then (\ref{distorder}) reduces to the Riemann-Liouville derivative $_{0}^{RL}D_{t}^{\beta}\left(f\left(t\right)\right)$. Note that the dimensions of $c(\alpha)$ are  
$\left[time\right]^{\alpha}/\left[length\right]^{2}$. 

In summary, we have just proved the following lemma:

\begin{lem}
If the relaxation modulus is given by $\ds G\left(t-t'\right)=\int_0^1 \frac{c(\alpha)}{\Gamma(1-\alpha)}(t-t')^{-\alpha}~d\alpha$, then the hereditary integral proposed by Boltzmann (\ref{sigma}), is given by:
\begin{equation}\label{distorder2}
\sigma\left(t\right)=_{0}^{RL}\mathbb{D}_{t}\gamma(t).
\end{equation}
\end{lem}



In this work we further develop the model and investigate its behaviour under deformation. We then derive a model with two elements of distributed order. We start by writing the strain as a function of stress and then we derive a more complex model that considers two elements of distributed order in series.
We obtain closed form solutions for simple weighting functions $c(\alpha)$ and present a detailed study of the behaviour of the model in simple flows. The work ends with some conclusions and plans for further investigation.


\section{Distributed Order Models}
\subsection{The distributed order element (distributed-springpot)}

\begin{figure}[!ht]
\centering
\begin{tabular}{cc}  \includegraphics[scale=0.13]{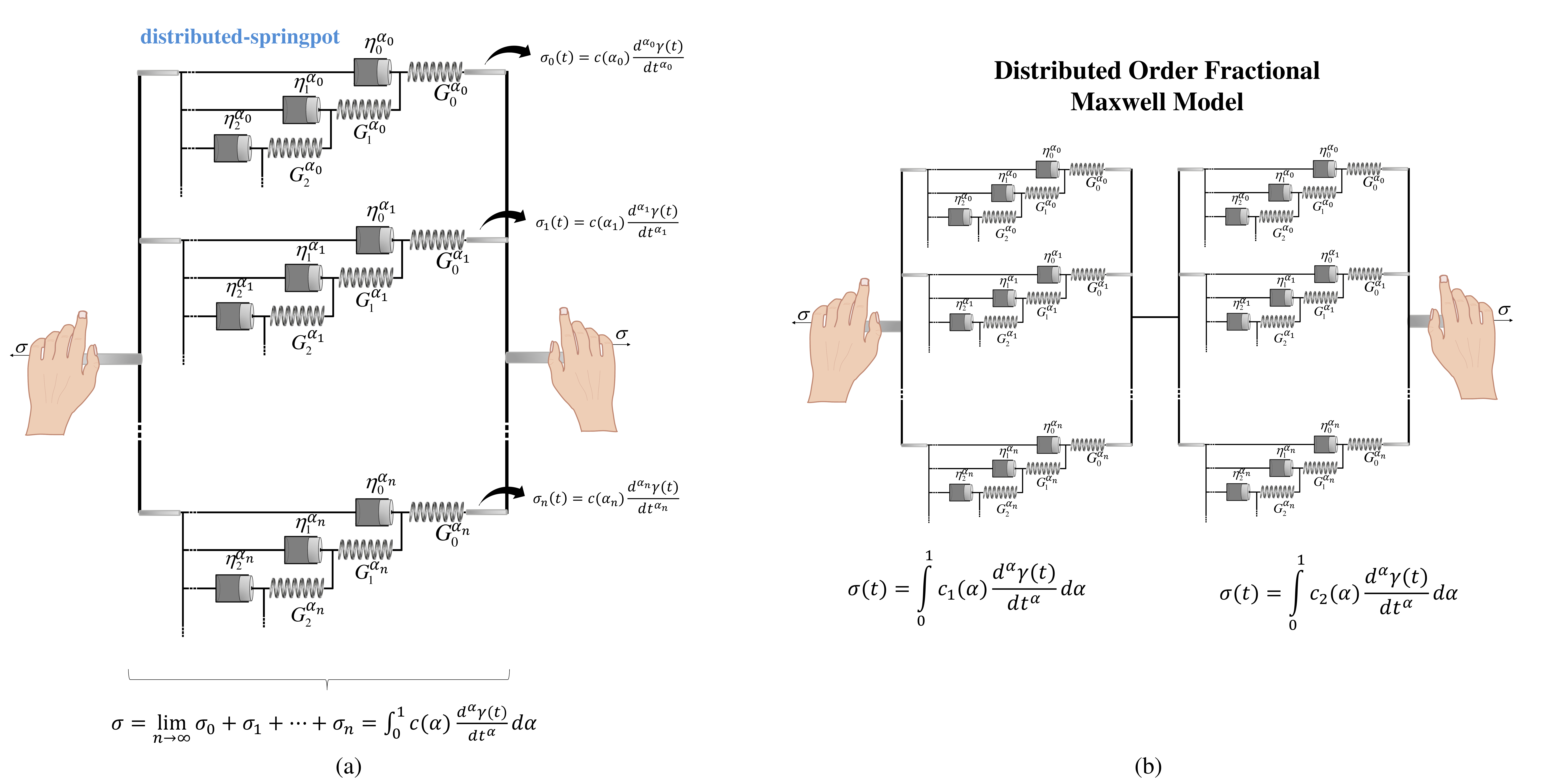}
\end{tabular}
\caption{(a) The distributed order element (distributed-springpot). (b) The Generalised Distribute-Order Maxwell Model (two distributed-springpots in series).}
\label{fig3}
\end{figure}

The distributed order model (\ref{DODEvisc}) can be represented by a weighted sum of springpots (see figure \ref{FMM}(a)), from now on referred to as \emph{distributed-springpot}. The mechanical analogue of a discrete version of the model (the weighted integral viewed as an infinite sum) is shown in figure \ref{fig3}.

For each springpot, the constants $\eta_i^{\alpha_j}$ and $G_i^{\alpha_j}$ ($i=0,1,...$ and $j=0,1,...$) vary (see \cite{Schiessel1993}) and also depend on the function $c(\alpha)$.
This arrangement of an infinite number of weighted springpots shows a more complex behaviour than the simple springpot. 

\begin{lem}\label{lemaprincipal}
The storage ($G'$) and loss (G'') moduli for the distributed order viscoelastic model  are given by:
\begin{eqnarray}
&&\label{storagemodulus}G'(\omega)=\int_0^1c(\alpha)\omega^{\alpha}\cos\left(\frac{\pi}{2}\alpha\right)~d
\alpha,\\
&&\label{lossmodulus}G''(\omega)=\int_0^1c(\alpha)\omega^{\alpha}\sin\left(\frac{\pi}{2}\alpha\right)~d
\alpha.
\end{eqnarray}
\end{lem}

\begin{pf} Let $\mathcal{L}\left(\cdot;s\right)$ be the Laplace transform, we have that,
\begin{equation}
\mathcal{L}\left(\sigma(t);s\right)=\tilde{\sigma}(s)=\int_0^1c(\alpha) \mathcal{L}\left( _{0}^{RL}D_{t}^{\alpha}\gamma(t);s\right)~d 
\alpha=\mathcal{L}\left(\gamma(t);s\right)\int_0^1c(\alpha)s^{\alpha}~d 
\alpha= \tilde{\gamma}(s)\int_0^1c(\alpha)s^{\alpha}~d 
\alpha.
\end{equation}
If we consider $s=i\omega$ and using the fact that $(i\omega)^{\alpha}={\omega}^{\alpha}\cos\left(\frac{\pi}{2}\alpha\right)+i\omega^{\alpha}\sin\left(\frac{\pi}{2}\alpha\right)$, we have:
\begin{equation}
G^*(i\omega)=\frac{\tilde\sigma(i\omega)}{\tilde\gamma(i\omega)}=\int_0^1c(\alpha)\omega^{\alpha}\cos\left(\frac{\pi}{2}\alpha\right)~d
\alpha+i\int_0^1c(\alpha)\omega^{\alpha}\sin\left(\frac{\pi}{2}\alpha\right)~d
\alpha
=G'(\omega)+iG''(\omega).
\end{equation}
\end{pf}

The storage and loss moduli will be used to obtain the relaxation modulus, $G(t)$ (the stress response to a jump in strain, $\gamma(t)=\gamma_0 H(t)$) and the creep compliance, $J(t)$ (the strain response to a jump in stress, $\sigma(t)=\sigma_0 H(t)$).

From \cite{Pipkin} we know that,

\begin{equation}\label{G1}
G(t)=\frac{2}{\pi} \int_{0}^{\infty}  \frac{G^{\prime}(\omega)}{\omega} \sin (\omega t) \mathrm{d} \omega=\mathcal{F}_{sin}^{-1}\left\{\frac{G^{\prime}(\omega)}{\omega} ; t\right\}
\end{equation}

and
\begin{equation}\label{G2}
    G(t)=\frac{2}{\pi} \int_{0}^{\infty} \frac{G^{\prime \prime}(\omega)}{\omega} \cos (\omega t) \mathrm{d} \omega=\mathcal{F}_{\mathrm{cos}}^{-1}\left\{\frac{G^{\prime \prime}(\omega)}{\omega} ; t\right\},
\end{equation}
with $\mathcal{F}_{sin}^{-1}$ and $\mathcal{F}_{cos}^{-1}$ the inverse Fourier \emph{sine} and \emph{cosine} transformations. 

For a distributed order element we have that,
$$
G(t)=\mathcal{F}_{sin}^{-1}\left\{\int_0^1c(\alpha)\omega^{\alpha-1}\cos\left(\frac{\pi}{2}\alpha\right)~d
\alpha ; t\right\}=\frac{2}{\pi} \int_{0}^{\infty} \int_0^1c(\alpha)\omega^{\alpha-1}\cos\left(\frac{\pi}{2}\alpha\right)~d\alpha \sin (\omega t) \mathrm{d} \omega.
$$
Note that from the above equality we have 
\begin{eqnarray}
\nonumber G(t)&=&\frac{2}{\pi} \int_0^1c(\alpha)\cos\left(\frac{\pi}{2}\alpha\right)\left( \int_{0}^{\infty}\omega^{\alpha-1}\sin (\omega t) ~d\omega\right)~d\alpha\\
\label{relaxmodulus2}&=&\frac{2}{\pi} \int_0^1c(\alpha)\cos\left(\frac{\pi}{2}\alpha\right)t^{-\alpha}
\Gamma(\alpha)\sin\left(\alpha \frac{\pi}{2}\right)\, d\alpha.
\end{eqnarray}
Using the identities $\ds \Gamma(\alpha)\Gamma(1-\alpha)=\frac{\pi}{\sin(\alpha \pi)},\, \forall \alpha \in (0,1)$, and $\ds 2\sin\left(\alpha \frac{\pi}{2}\right)\cos\left(\alpha \frac{\pi}{2}\right)=\sin(\alpha \pi)$ \cite{Mathai}, from (\ref{relaxmodulus2}) we obtain, as expected, equation (\ref{relaxd}),  $$\ds G(t)= \int_0^1\frac{c(\alpha)}{\Gamma(1-\alpha)}t^{-\alpha}d\alpha .$$

The complex compliance is defined by  $J^{*}(\omega)=1 / G^{*}(\omega)$, with the storage and loss compliances given by $J^{\prime}(\omega)=$ $\operatorname{Re} J^{*}(\omega)$ and $J^{\prime \prime}(\omega)=-\operatorname{Im} J^{*}(\omega)$, respectively:

$$
J^{\prime}(\omega)=\frac{\int_0^1 c(\alpha)\omega^{\alpha} \cos \left(\frac{\pi  \alpha}{2}\right) \, d\alpha}{\left(\int_0^1 c(\alpha)\omega^{\alpha} \cos \left(\frac{\pi  \alpha}{2}\right) \, d\alpha\right)^2 + \left(\int_0^1 c(\alpha)\omega^{\alpha} \sin \left(\frac{\pi  \alpha}{2}\right) \, d\alpha\right)^2},
$$

$$
J^{\prime \prime}(\omega)=\frac{\int_0^1 c(\alpha)\omega^{\alpha} \sin \left(\frac{\pi  \alpha}{2}\right) \, d\alpha}{\left(\int_0^1 c(\alpha)\omega^{\alpha} \cos \left(\frac{\pi  \alpha}{2}\right) \, d\alpha\right)^2 + \left(\int_0^1 c(\alpha)\omega^{\alpha} \sin \left(\frac{\pi  \alpha}{2}\right) \, d\alpha\right)^2}.
$$

The creep compliance is given by:
\begin{equation}
    \label{CreepComp}\frac{\mathrm{d} J(t)}{\mathrm{d} t}=\mathcal{F}_{\mathrm{cos}}^{-1}\left\{J^{\prime}(\omega) ; t\right\}=-\mathcal{F}_{sin}^{-1}\left\{J^{\prime \prime}(\omega) ; t\right\},
\end{equation}
that is,


\begin{eqnarray}
\nonumber \frac{\mathrm{d} J(t)}{\mathrm{d} t} &=&\mathcal{F}_{\mathrm{cos}}^{-1}\left\{J^{\prime}; t\right\} = F(t)\\
\label{compliance}&=&\frac{2}{\pi} \int_{0}^{\infty} \left(\frac{\int_0^1 c(\alpha)\omega^{\alpha} \cos \left(\frac{\pi  \alpha}{2}\right) \, d\alpha}{\left(\int_0^1 c(\alpha)\omega^{\alpha} \cos \left(\frac{\pi  \alpha}{2}\right) \, d\alpha\right)^2 + \left(\int_0^1 c(\alpha)\omega^{\alpha} \sin \left(\frac{\pi  \alpha}{2}\right) \, d\alpha\right)^2}\right) \cos (\omega t) \mathrm{d} \omega,
\end{eqnarray}

Integrating and using the fact that \cite{Schiessel1995}
$$\lim_{t\rightarrow 0} J(t)= \lim_{t\rightarrow 0} 1 / G(t)=0,$$ we have
$$
J(t)=\lim_{t^{\star}\rightarrow 0} \int_{t^{\star}}^tF(s)ds,
$$
where $F(\cdot)$ is given by (\ref{compliance}).

It is difficult to obtain the general analytical expression of the compliance due to the complexity of the involved integrals. Nevertheless, as we show next, for particular cases of $c(\alpha)$ the analytical solution is possible.

\begin{itemize}
    \item Single-order case: $c(\alpha)=\delta(\alpha-\beta), \alpha<\beta<1$
    
We have that 
    $$ \frac{\mathrm{d} J(t)}{\mathrm{d} t}=\frac{t^{\beta-1}}{\Gamma(\beta)}
    $$ 
    resulting in
         $$ J(t)=\frac{t^{\beta}}{\Gamma(1+\beta)}.$$ 
Note that this particular case of creep compliance is the same as the one presented in \cite{Schiessel1995} for a much simpler model (see their equation 21). This reinforces that fact that our model generalises the classical and fractional models.

    \item Uniformly distributed-order case: $\displaystyle c(\alpha)=1$
    
    $$
    J(t)=    \frac{\sqrt{\pi }}{t} G_{1,3}^{2,1}\left(\frac{t^2}{4}\bigg\rvert
    \begin{array}{c}
     1 \\
     1,1,\frac{1}{2} \\
    \end{array}
    \right)-\text{Chi}(t)\cosh (t)+\text{Shi}(t) \sinh (t)+\ln (t)+E_\gamma,
    $$ 
where $E_\gamma$ is the Euler's constant ($E_\gamma\sim 0,57721566490153286$), $G()$ is the Meijer G-function. $\text{Chi}(t)$ and $\text{Shi}(t)$ are the hyperbolic cosine and the hyperbolic sine integrals, respectively \cite{Mathai,Meijer}.

    \item Case: $\displaystyle c(\alpha)=e^\alpha$
    $$J(t)=e^{t/e} \left(\text{Shi}\left(\frac{t}{e}\right)-\text{Chi}\left(\frac{t}{e}\right)\right)+\ln (t) +E_\gamma -1.$$
\end{itemize}

\begin{figure}[!ht]
\centering
\begin{tabular}{cc}  \includegraphics[scale=0.60]{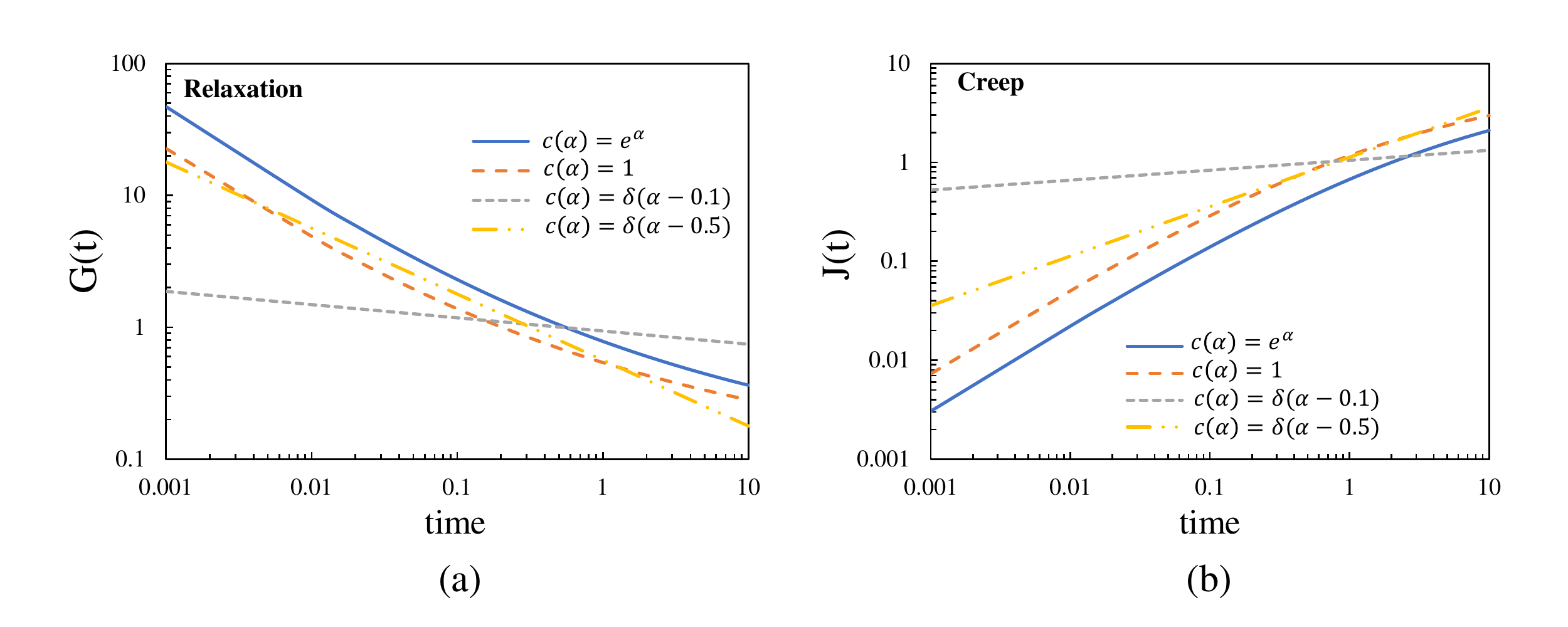}
\end{tabular}
\caption{(a) Relaxation; (b) Creep. Time is given in seconds. Plots of the relaxation modulus and creep compliance obtained for the distributed order model given by equation \ref{distorder2}. Four weighting functions are considered: $c(\alpha)=e^{\alpha};~c(\alpha)=1;~c(\alpha)=\delta(\alpha-0.1);~c(\alpha)=\delta(\alpha-0.5)$.}
\label{relCreep}
\end{figure}

The behaviour of these relaxation and creep functions (for the different weighting functions considered) is shown in figure \ref{relCreep}.

The results are compared with the special case of the relaxation modulus and creep compliance obtained for a simple springpot (see figure \ref{FMM}(a)). It can be seen that the springpot exhibits different behaviour compared to its distributed-order version. This is due to the fact that in the distributed-springpot we have considered distributions that contain all the $\alpha$-order derivatives $\in [0,1]$ with similar weights. This way, the individual influence of each $\alpha$-order derivative is mitigated. The fact that we can choose any weighting function gives the model a high degree of freedom to fit experimental data obtained from different material such as polymer melts, soft gels, elastic solids, etc.

\noindent\textbf{Remark:} The weighting functions could be further generalised by introducing extra parameters. For example, instead of $c(\alpha)=e^{\alpha}$ one could have used $c(\alpha)=e^{a\alpha}$ with $a\in\mathbb{R}$, or, $c(\alpha)=a$, instead of $c(\alpha)=1$. The reason for not doing this is to keep the paper as simple as possible. These generalisations can be easily introduced by the interested reader.


\section{The Generalised Distributed-Order Maxwell Model}
\noindent

As in the classical (fractional) Maxwell model, where we have a spring and a dashpot (two springpots), in series, to derive the Generalised Distributed-Order Maxwell Model (GDOMM) we have to connect two distributed-springpots in series (see figure \ref{fig3}(b)).

The distributed order model (\ref{DODEvisc}) defines the stress $(\sigma(t))$ as a function of strain $(\gamma(t))$. To obtain an analytical expression for the stress of two distributed springpots connected in series, it would be useful to define the strain $(\gamma(t))$ as a function of the stress $(\sigma(t))$ for both elements. As we will see next, a closed-form expression for this can be obtained. Due to the complexity of the involved operators such expression is difficult to handle, although it can be easily derived for certain choices of the weighting function $c(\alpha)$.


\subsection{Strain $(\gamma(t))$ as a function of stress $(\sigma(t))$}

In what follows $\displaystyle\tilde{f}(s)$ denotes the Laplace transform of $f(t)$, that is 
\begin{eqnarray}
\nonumber && \mathcal{L}\left(f(t);s\right)= \tilde{f}(s)\\
\nonumber && \mathcal{L}^{-1}\left(\tilde{f}(s);t\right)= f(t).
\end{eqnarray}

\begin{thm}\label{teorema2}
The strain $\gamma(t)$ that satisfies the distributed order equation (\ref{distorder2}) is described as a function of the stress, $\sigma(t)$, through the following equality
\begin{equation}\label{defgamma}
\gamma(t)=\sigma(t)\ast F(t),  
\end{equation} where  $\ast$ denotes the convolution of $\sigma(t)$ and $F(t)$. The function $F(t)$ is defined by  
\begin{equation}
    F(t)=\frac{1}{\pi} \int_{0}^{+\infty} e^{-r t} \, \, \frac{\sin (\theta\pi)}{\rho} dr, \label{defF}
\end{equation} with $\rho$ and $\theta$ given by $\rho \equiv \rho(r)=\left|b\left(r e^{i \pi}\right)\right|$ and $\theta\equiv \theta(r)=\frac{1}{\pi} \operatorname{Arg}\left|b\left(r e^{i \pi}\right)\right|$, respectively, and $b(\cdot)$ is defined as $b(s)=\int_0^1 c(\alpha)s^{\alpha}~d\alpha$.
\end{thm}
\begin{pf}
Consider the distributed order fractional viscoleastic model given by
\begin{equation}\label{eqDO1}
    \sigma\left(t\right)=_{0}^{RL}\mathbb{D}_{t}\left(\gamma(t)\right).
\end{equation}
By using Laplace transform in (\ref{eqDO1}) we have 
\begin{equation}\label{eqDO12}
\mathcal{L}\left(\sigma\left(t\right);s\right)=\int_0^1 c(\alpha) \mathcal{L}\left(
_{0}^{RL}D_{t}^{\alpha}(\gamma(t));s\right) d\alpha.
\end{equation}.

The Laplace transform of the $\alpha-th$   derivative, in the Riemman Liouville sense, $\displaystyle 0<\alpha <1$, is given by (\cite{Mainardi97})
$$\mathcal{L}\left(_{0}^{RL}D_{t}^{\alpha}(\gamma(t));s
\right)=s^{\alpha}\mathcal{L}\left(\gamma(t);s
\right)$$ then 
\begin{eqnarray}\label{eqcontas2}
\mathcal{L}\left(\sigma\left(t\right);s\right)&=&
\int_0^1 c(\alpha)s^{\alpha}
\mathcal{L}\left(\gamma(t);s\right)d\alpha=b(s)
\mathcal{L}\left(\gamma(t);s\right)d\alpha,\end{eqnarray} where $$b(s)=\int_0^1 c(\alpha)s^{\alpha}~d\alpha.$$
From (\ref{eqcontas2}) we have 
\begin{equation}\label{Lapla}
    \tilde{\gamma}\left(s\right)=\frac{\tilde{\sigma}(s)}{b(s)}.\end{equation}
Applying the inverse of the Laplace transform to both sides of the previous equation, we obtain:
\begin{equation}
\gamma(t)= \mathcal{L}^{-1}\left(\frac{\tilde{\sigma}(s)}{b(s)};t\right)= \sigma(t)\ast \mathcal{L}^{-1}\left(\frac{1}{b(s)};t\right). \label{eq1}
\end{equation}
\bigskip

Let us now analyse $\mathcal{L}^{-1}\left(\frac{1}{b(s)} ; t\right)$. 

Taking the Titchmarsh theorem into account:
$$
\mathcal{L}^{-1}\left(\frac{1}{b(s)} ; t\right)=-\frac{1}{\pi} \int_{0}^{+\infty} e^{-r t} \operatorname{Im}\left(\frac{1}{b\left(r e^{i \pi)}\right)}\right) dr.
$$
In order to simplify the expression above, we write:

\begin{eqnarray}\label{aux}
b\left(r e^{i \pi}\right)=\rho \cos (\theta \pi)+i \rho \operatorname{sin}(\theta \pi) ,\end{eqnarray} where
\begin{eqnarray}
&&\label{defrho} \rho \equiv \rho(r)=\left|b\left(r e^{i \pi}\right)\right|,\\
&& \label{deftheta} \theta\equiv \theta(r)=\frac{1}{\pi} \operatorname{Arg}\left|b\left(r e^{i \pi}\right)\right|.
\end{eqnarray}

Then:
\begin{eqnarray}
\nonumber 
 -\mathrm{Im}\left(\frac{1}{b\left(r e^{i \pi}\right)}\right)  &=&\displaystyle-\operatorname{Im}\left(\frac{1}{\rho \cos (\theta \pi)+i \rho \operatorname{sin}(\theta \pi)}\right)=-\operatorname{Im}\left(\frac{\rho \cos (\theta \pi)-i \rho \sin (\theta  \pi)}{\rho^{2}}\right)\\
 \nonumber &=&\frac{\operatorname{sin}(\theta  \pi)}{\rho}.
\end{eqnarray}

Therefore,
\begin{equation}
  \mathcal{L}^{-1}\left(\frac{1}{b(s)} ; t\right)= F(t)=
  \frac{1}{\pi} \int_{0}^{+\infty} e^{-r t} \, \, \frac{\sin (\theta  \pi)}{\rho} dr. \label{aux1}
\end{equation}
Thus, from (\ref{eq1}) we obtain
\begin{equation}\label{defgamma}
\gamma(t)=\sigma(t)\ast F(t), 
\end{equation} where $F$ is defined by (\ref{defF}).
\end{pf}

\vspace*{0.5cm}


Next, we consider some particular cases for the function $c(\alpha)$. Using Theorem \ref{teorema2} we obtain the expression of strain as function of stress. The details can be found in Appendix A.

\begin{itemize}
    \item Single-order case: $c(\alpha)=\delta(\alpha-\beta),\,  0<\alpha<\beta<1$.
    \begin{equation}\label{expgammacase1}\gamma(t)=\sigma(t)\ast \frac{t^{\beta-1}}{\Gamma(\beta)}=
    \frac{1}{\Gamma(\beta)}\int_0^t (t-s)^{\beta-1}\sigma(s)ds,\end{equation}
 This means that the strain is equal to the Riemann-Liouville integral (of order $\beta$) of $\sigma(t)$. 
 Once again, this particular case of our model is in accordance with the results presented in \cite{Schiessel1995} for a simpler model.

    \item Uniformly distributed-order case: $\displaystyle c(\alpha)=1,\quad 0<\alpha<1$.
    
    \begin{equation}\label{gammacaso2}
    \gamma(t)=\int_{0}^{t} \sigma(s) e^{t-s} \Gamma(0, t-s) ds,
    \end{equation}
    where $\ds \Gamma(0, t)$ is the incomplete Gamma function given by 
    \begin{equation}
    \label{GamaFunc}    \Gamma(0, t)=\int_{t}^{+\infty} t^{-1} e^{-t} d t.  
    \end{equation}
  
    \item Exponential case:  $\displaystyle c(\alpha)=e^{\alpha},\quad 0<\alpha<1.$
    
    \begin{equation}\label{gammacasoexp}
    \gamma(t)=\int_{0}^{t} \sigma(s) e^{\frac{t-s}{e}-1} \Gamma\left(0, \frac{t-s}{e}\right) ds,
    \end{equation}
    where $\ds \Gamma(0, t)$ is the incomplete Gamma function given by (\ref{GamaFunc}).

\end{itemize}

\subsection{Two Distributed Order Elements in Series}
\noindent

In this section, we intend, using Theorem \ref{teorema2}, to generalise the Distributed Order Model (equation \ref{distorder2}) by considering two distributed order elements (distributed-springpots) in series. This will result in the Generalised Distributed-Order Maxwell Model.

We consider two distributed-order elements in series, and that the stress is the same for both elements (see Figure \ref{fig3} (b)). From Theorem \ref{teorema2}, the respective stress-strain relationships for two elements are given by 
\begin{eqnarray}
&&\label{element1} \gamma_1(t)=\sigma(t)\ast \mathcal{L}^{-1}\left(\frac{1}{b_1(s)};t\right),\\
&&\label{element2} \gamma_2(t)=\sigma(t)\ast \mathcal{L}^{-1}\left(\frac{1}{b_2(s)};t\right),
\end{eqnarray}
where $\ds b_i(s)=\int_0^1c_i(\alpha)s^{\alpha}ds,$ and the functions $c_i,\, i=1,2, $ correspond to the weighting functions related with the distributed derivative. As in the classical case, we have that the total deformation is given by the sum of the deformations obtained for the two elements in series ($\ds \gamma(t)=\gamma_1(t)+\gamma_2(t)$). From (\ref{element1}) and (\ref{element2}) it follows
\begin{eqnarray}
\nonumber \gamma(t)&=&\sigma(t)\ast \mathcal{L}^{-1}\left(\frac{1}{b_1(s)}+\frac{1}{b_2(s)};t\right),\\
\nonumber &=&\int_0^t \sigma(t-s) \mathcal{L}^{-1}\left(\frac{1}{b_1(s)}+\frac{1}{b_2(s)};t\right)ds\\
\label{elementoUNICO}&=&
\int_0^t \sigma(t-s)\left( \mathcal{L}^{-1}\left(\frac{1}{b_1(s)};t\right)+\mathcal{L}^{-1}\left(\frac{1}{b_2(s)};t\right)\right)ds.
\end{eqnarray}

Using the same arguments as in Theorem \ref{teorema2}, we have 

\begin{eqnarray}
\label{elementoUNICO2}\gamma(t)=\frac{1}{\pi}\int_0^t \sigma(t-s)\int_0^{+\infty}  e^{-r s}\left(\frac{\sin (\theta_1  \pi)}{\rho_1}+\frac{\sin (\theta_2  \pi)}{\rho_2}\right)dr \, ds,\end{eqnarray} where 
$\rho_k,\, \theta_k,\quad k=1,2$, are such that $$b_k(re^{i\pi})=\rho_k\cos(\theta_k \pi)+i\rho_k\sin(\theta_k \pi),\, k=1,2.$$

\medskip

On the other hand, if we apply the Laplace transform to both sides of (\ref{elementoUNICO}) it follows
\begin{eqnarray}
\nonumber \mathcal{L}\left(\gamma(t);s\right)=\tilde{\gamma}(s)&=&\mathcal{L}\left(\sigma(t)\ast \mathcal{L}^{-1}\left(\frac{1}{b_1(s)}+\frac{1}{b_2(s)};t\right);s\right)\\
\label{igualdadeusar} &=& \mathcal{L}\left(\sigma(t);s\right)\left(\frac{1}{b_1(s)}+\frac{1}{b_2(s)}\right)=
\tilde{\sigma}(s)\left(\frac{1}{b_1(s)}+\frac{1}{b_2(s)}\right)
\end{eqnarray}
and we have 
\begin{equation}
\tilde{\sigma}(s)=\frac{b_1(s)b_2(s)}{b_1(s)+b_2(s)}\tilde{\gamma}(s) \label{igualdadeusar}.
\end{equation}
Applying the inverse Laplace transform to both sides of the above equation follows 
\begin{eqnarray}
\nonumber \sigma(t)&=&\mathcal{L}^{-1}\left(
\frac{b_1(s)b_2(s)}{b_1(s)+b_2(s)}\tilde{\gamma}(s);t\right)\\
\nonumber &=&\mathcal{L}^{-1}\left(\frac{b_1(s)b_2(s)}{b_1(s)+b_2(s)};t\right)\ast \mathcal{L}^{-1}\left(\tilde{\gamma}(s);t\right)\\
\label{gamma2ele}&=& \int_0^t \gamma(t-y)\mathcal{L}^{-1}\left(\frac{b_1(s)b_2(s)}{b_1(s)+b_2(s)};y\right)dy.
\end{eqnarray}

The Generalised Distributed-Order Maxwell Model is then given by:
\begin{equation}
\sigma(t)= \int_0^t \gamma(t-y)\mathcal{L}^{-1}\left(\frac{b_1(s)b_2(s)}{b_1(s)+b_2(s)};y\right)dy \label{gamma2ele}.
\end{equation}

\medskip

In order to obtain the complex modulus 
we can use equation (\ref{igualdadeusar}) and consider $s=i\omega$ leading to the complex modulus
\begin{equation}\label{GStar}G^*(\omega)= \frac{\tilde{\sigma}(i\omega)}{\tilde{\gamma}(i\omega)}=\frac{b_1(i\omega)b_2(i\omega)}{b_1(i\omega)+b_2(i\omega)}.\end{equation}
In order to obtain the storage and loss modulus for the Generalised Distributed-Order Maxwell Model, we must proceed as in Lemma \ref{lemaprincipal}. After some calculations, from the expression of the complex modulus, we can obtain the storage and loss modulus, which are given by
\begin{eqnarray}
\label{StorageMod} &&G'(w)=Re(G^*(w))=\frac{C_1(C_2^2+S_2^2) +C_2(C_1^2+S_1^2)}{(C_1+C_2)^2+(S_1+S_2)^2 },\\
\label{LossMod} &&G''(w)=Im(G^*(w))=\frac{S_2(C_1^2+S_1^2) +S_1(C_2^2+S_2^2)}{(C_1+C_2)^2+(S_1+S_2)^2  },
\end{eqnarray}
where 
\begin{eqnarray}
\label{bc} && C_i=\int_0^1 c_i(\alpha)w^\alpha \cos\left(\frac{\pi}{2}\alpha\right)d\alpha,\, i=1,2,\\
\label{bs} &&S_i=\int_0^1 c_i(\alpha)w^\alpha \sin\left(\frac{\pi}{2}\alpha\right)d\alpha,\, i=1,2,.
\end{eqnarray}

The complex compliance satisfies $\displaystyle{J^*(w) = \frac{1}{G^*(w)}}$.  The storage and loss compliances are given by $J'(w) =
Re(J^*(w))$ and $J''(w) = -Im(J^*(w)$, respectively. Therefore, from (\ref{GStar}), together with some algebra, it follows that
\begin{eqnarray}
\label{StorageCOMP} &&J'(w)=\frac{C_1(C_2^2+S_2^2) +C_2(C_1^2+S_1^2)}{(C_1C_2-S_1S_2)^2+(C_1S_2+S_1C_2)^2 },\\
\label{LossCOMP} &&J''(w)=\frac{S_2(C_1^2+S_1^2) +S_1(C_2^2+S_2^2)}{(C_1C_2-S_1S_2)^2+(C_1S_2+S_1C_2)^2 },
\end{eqnarray}
where {\color{black}${C}_i,\, {S}_i,~ (i=1,2)$} are given by (\ref{bc}) and (\ref{bs}), respectively.

\smallskip

\noindent\textbf{Remark:} Constitutive equations must satisfy the restrictions that follow from the second law of thermodynamics. These restrictions require that the tangent of the mechanical loss angle is non-negative, $\tan(\delta)>0$, and that both $G'$ and $G''$ are positive for all values of $w$.  \\
The later restriction is easily verified, from (\ref{StorageMod}) and (\ref{LossMod}), and the tangent of the mechanical loss angle for the generalized distributed order Maxwell model is given by 
\begin{equation}
\label{tanloss} \tan(\delta)=\frac{S_2(C_1^2+S_1^2) +S_1(C_2^2+S_2^2)}{C_1(C_2^2+S_2^2) +C_2(C_1^2+S_1^2)}.    
\end{equation}
Note that these expressions/restrictions are $c(\alpha)$ dependent.

\bigskip

We will now derive the expressions for the storage and loss modulus, relaxation modulus and creep compliance for particular cases of $c(\alpha)$.
\begin{itemize}
    \item Generalised-Distributed Order Maxwell Model with $c_1(\alpha)=\delta(\alpha-\beta_1)$ and $c_2(\alpha)=\delta(\alpha-\beta_2)$, where $\beta_1,\, \beta_2\in (0,1)$ and, it is assumed, without loss of generality, that $\beta_1\geq \beta_2.$

   In this case $b_1(s)=s^{\beta_1}$ and $b_2(s)=s^{\beta_2}$.  Therefore 
   $$G^*(w)=\frac{(iw)^{\beta_1}}{1+(iw)^{\beta_1-\beta_2}}$$ and from (\ref{StorageMod}) and (\ref{LossMod}) we have 
   \begin{eqnarray}
\label{StorageModexemp1} &&G'(w)=\frac{w^{\beta_1}\cos(\frac{\beta_1\pi}{2})+w^{2\beta_1-\beta_2}\cos(\frac{\beta_2\pi}{2})}{1+w^{2(\beta_1-\beta_2)}+2w^{\beta_1-\beta_2}\cos(\frac{\pi}{2}(\beta_1-\beta_2))},
\\
\label{LossModexemp1} &&G''(w)=\frac{w^{\beta_1}\sin(\frac{\beta_1\pi}{2})+w^{2\beta_1-\beta_2}\sin(\frac{\beta_2\pi}{2})}{1+w^{2(\beta_1-\beta_2)}+2w^{\beta_1-\beta_2}\cos(\frac{\pi}{2}(\beta_1-\beta_2))}.
\end{eqnarray}
   The expressions above are in agreement with the results presented in \cite{Schiessel1995}.
 Following the same arguments as in \cite{Schiessel1995}, the Mellin transform of $\displaystyle{\frac{G''(w)}{w}}$ is given by
   \begin{equation}\label{ig1}
       \mathcal{M}\left(\frac{G''(w)}{w};z\right)= \frac{\pi}{\beta_1-\beta_2}
       \frac{\cos(z \,\pi/2)}{\sin\left(\pi\frac{z+\beta_1-1}{\beta_1-\beta_2}\right)},  \end{equation} and from (\ref{G2}) it follows that
   \begin{equation}\label{ig2}
      \mathcal{M}\left(\frac{G''(w)}{w};z\right)=
      \mathcal{M}\left(\mathcal{F}_{\mathrm{cos}}\left(G(t);w\right);z\right)= \Gamma(z)\cos(\frac{\pi}{2}z)\mathcal{M}\left(G(t);1-z\right).
   \end{equation}
   Therefore, from (\ref{ig1}) and (\ref{ig2}) we have 
    \begin{equation}\label{ig3}
     \mathcal{M}\left(G(t);1-z\right)=\frac{\pi}{\Gamma(z)(\beta_1-\beta_2)}\left(\sin\left(\frac{\pi}{\beta_1-\beta_2}(z+\beta_1-1)\right)\right)^{-1}.
   \end{equation} 

   Using the identity  $\ds \Gamma(\alpha)\Gamma(1-\alpha)=\frac{\pi}{\sin(\alpha \pi)},\, \forall \alpha \in (0,1)$, the previous equality leads to the following expression for the relaxation modulus, that depends on the inverse of the Mellin transform,
    \begin{equation}\label{ig4}
     G(t)=\frac{1}{\beta_1-\beta_2}\mathcal{M}^{-1}\left(
     \frac{\Gamma(\frac{\beta_1-z}{\beta_1-\beta_2})\Gamma(\frac{z-\beta_2}{\beta_1-\beta_2})}{\Gamma(1-z)}
     \right).\end{equation}
     
    Following \cite{Schiessel1995} and using the relationship between the inverse Mellin transform and the definition of the Fox H-function (\cite{HFunction}) we obtain the final expression for the relaxation modulus,
    \begin{eqnarray}
    \label{RMexemp1}G(t)=t^{-\beta_2}E_{\beta_1-\beta_2,1-\beta_2}(-t^{\beta_1-\beta_2}), 
    \end{eqnarray} where $E_{\beta_1-\beta_2,1-\beta_2}(\, )$ is the generalized Mittag-Leffler function.
    
    \vspace*{0.5cm}
    The complex compliance is given by 
    \begin{eqnarray}
   \nonumber  J^{*}(w)&=&\frac{1}{G^*(w)}=(iw)^{-\beta_1}+(iw)^{-\beta_2}\\
       \nonumber&=& w^{-\beta_1}\cos\left(-\frac{\pi}{2}\beta_1\right)+w^{-\beta_2}\cos\left(-\frac{\pi}{2}\beta_2\right)+i\left(w^{-\beta_1}\sin\left(-\frac{\pi}{2}\beta_1\right)+w^{-\beta_2}\sin\left(-\frac{\pi}{2}\beta_2\right)\right).\\
        \label{compilance}
    \end{eqnarray}
    
    From (\ref{StorageCOMP}) and (\ref{LossCOMP}) the storage and loss compliance's are given by
    \begin{eqnarray}
    \label{Compig1}J'(w)=w^{-\beta_1}\cos\left(\frac{\pi}{2}\beta_1\right)+w^{-\beta_2}\cos\left(\frac{\pi}{2}\beta_2\right),\\
    \label{Lossig1}J''(w)=w^{-\beta_1}\sin\left(\frac{\pi}{2}\beta_1\right)+w^{-\beta_2}\sin\left(\frac{\pi}{2}\beta_2\right),
    \end{eqnarray}
    respectively. These results are in agreement with the results obtained by \cite{Schiessel1995}. Using the definition of creep compliance, the linearity of the operator $\mathcal{F}_{\mathrm{cos}}^{-1}$ and equality (21) of  \cite{Schiessel1995}, we obtain a simple analytical expression for the creep compliance: 
    \begin{equation}
        \label{Creep} J(t)=\frac{t^{\beta_1}}{\Gamma(1+\beta_1)}+
        \frac{t^{\beta_2}}{\Gamma(1+\beta_2)}.
    \end{equation}

    \item Generalised Distributed-Order Maxwell Model with $c_1(\alpha)=\delta(\alpha-\beta),\, \beta \in (0,1),$ and $c_2(\alpha)=1$.

   In this case $b_1(s)=s^{\beta}$ and $b_2(s)=\frac{s-1}{\ln(s)}$.  Therefore 
   $$G^*(w)=\frac{iw-1}{\ln(iw)+(iw)^{-\beta}(iw-1)}$$ and from (\ref{StorageMod}) and (\ref{LossMod}) we have 
     \begin{eqnarray}
\label{StorageModexemp2} &&G'(w)=\frac{ 2w^{2\beta}g_1(w)+4 w^{\beta} \left(w^2+1\right) \cos \left(\frac{\pi  \beta}{2}\right)}{ w^{2 \beta}(\pi ^2+4 \ln^2(w))+4w^{\beta}\left( \sin \left(\frac{\pi  \beta}{2}\right) g_2(w)+ 
\cos \left(\frac{\pi  \beta}{2}\right)g_1(w)\right)+4 w^2+4},
\\
\label{LossModexemp2} &&G''(w)=
\frac{ 2w^{2\beta} g_2(w)+4 w^{\beta}  \left(w^2+1\right) \sin \left(\frac{\pi  \beta}{2}\right)}
{ w^{2 \beta}(\pi ^2+4 \ln ^2(w))+4w^{\beta}\left( \sin \left(\frac{\pi  \beta}{2}\right) g_2(w)+ 
\cos \left(\frac{\pi  \beta}{2}\right) g_1(w)\right)+4 w^2+4},
\end{eqnarray}
  where $g_1(w)=\pi  w-2 \ln (w)$ and $g_2(w)=2 w \ln (w)+\pi)$.
  
  In this case is not possible to explicitly obtain the expression of the relaxation modulus that will be given by 
  \begin{equation}
      \label{RModulusExamp2}G(t)=\mathcal{F}_{\mathrm{cos}}^{-1}\left\{\frac{G^{\prime \prime}(\omega)}{\omega} ; t\right\},
  \end{equation}
  where $G''$ is given by (\ref{LossModexemp2}).
  However, it is possible obtain an approximation of $G(t)$ using an appropriate numerical method to approximate the integral that defines $G(t)$. 
    \vspace*{0.5cm}
    
   On the other hand, it is possible compute the creep compliance. Indeed,  the complex compliance is given by
    \begin{equation}
     \label{compilanceexamp2} J^{*}(w)=\frac{1}{G^*(w)}=\frac{\ln(iw)}{iw-1}+(iw)^{-\beta}.
    \end{equation}
    
    From (\ref{StorageCOMP}) and (\ref{LossCOMP}) the storage and loss compliance's are given by
    \begin{eqnarray}
    \label{Compig2}J'(w)=
     \frac{\pi  w-2 \log (w)}{2 \left(w^2+1\right)}+w^{-\beta}\cos \left(\frac{\pi  \beta}{2}\right),\\
    \label{Lossig2}J''(w)=\frac{\pi +2 w\log (w)}{2 \left(w^2+1\right)}+w^{-\beta}\sin \left(\frac{\pi  \beta}{2}\right),
    \end{eqnarray}
    respectively.
    
Finally, from (\ref{CreepComp}) it follows that the creep compliance is given by 
 \begin{eqnarray}
 \nonumber J(t) &=&
 \frac{2}{\pi\beta}t^{\beta}\Gamma(1-\beta)\sin\left(\beta \frac{\pi}{2}\right)\cos\left(\beta \frac{\pi}{2}\right)\\
 \nonumber &+&
 \frac{\sqrt{\pi } t}{4} G_{1,3}^{2,1}\left(\frac{t^2}{4}\bigg\rvert
    \begin{array}{c}
     0 \\
     0,0,-\frac{1}{2} \\
    \end{array}
    \right)-\text{Chi}(t)\cosh (t)+\text{Shi}(t) \sinh (t)+\ln (t)+E_\gamma,
 \end{eqnarray}      
where $E_\gamma$ is the Euler's constant, $G_{1,3}^{2,1}(\cdot)$ is the Meijer G-function. $\text{Chi}(t)$ and $\text{Shi}(t)$ are the hyperbolic cosine and the hyperbolic sine integrals, respectively.

       \item Generalised Distributed-Order Maxwell Model with $c_1(\alpha)=\delta(\alpha-\beta),\, \beta \in (0,1)$,  and $c_2(\alpha)=e^{\alpha}$.

   In this case $b_1(s)=s^{\beta}$ and $b_2(s)=\frac{es-1}{1+\ln(s)}$.  Therefore 
   $$G^*(w)=\frac{e(iw)-1}{1+\ln(iw)+(iw)^{-\beta}(eiw-1)}$$ and from (\ref{StorageMod}) and (\ref{LossMod}) we have 
     \begin{eqnarray}
\nonumber &&G'(w)=\frac{2g_2(w)+4 w^{-\beta}\left(e^2 w^2+1\right) 
 \sin\left(\frac{\pi  }{2}\beta\right)}
 {\pi ^2+ 4+4\ln^2(w)+8 \ln(w)+4 w^{-\beta}\left( \sin \left(\frac{\pi  \beta}{2}\right) g_2(w)+\cos \left(\frac{\pi }{2} \beta\right) g_1(w)\right)+g_3(w)},\\
 \label{StorageModexemp3}
\\
\nonumber &&G''(w)=
\frac{2g_1(w)+4 w^{-\beta}\left(e^2 w^2+1\right) 
 \cos\left(\frac{\pi  }{2}\beta\right)}
 {\pi ^2+ 4+4\ln^2(w)+8 \ln(w)+4 w^{-\beta}\left( \sin \left(\frac{\pi  \beta}{2}\right) g_2(w)+\cos \left(\frac{\pi }{2} \beta\right) g_1(w)\right)+g_3(w)}
,\\
\label{LossModexemp3}
\end{eqnarray}
  where
  \begin{eqnarray}
  \label{g1exemplo3}&&g_1(w)=e \pi w-2\ln(w)-2,\\
  \label{g2exemplo3}&&g_2(w)=2 e w+2 e w \ln (w)+\pi,\\
   \label{g3exemplo3}&&g_3(w)= 4e^2w^{2-2\beta}+4w^{-2\beta}.
  \end{eqnarray}
  
Again, in this case is not possible to obtain, analytically, the expression of the relaxation modulus, given by 
  \begin{equation}
      \label{RModulusExamp3}G(t)=\mathcal{F}_{\mathrm{cos}}^{-1}\left\{\frac{G^{\prime \prime}(\omega)}{\omega} ; t\right\},
  \end{equation}
  where $\ds G^{\prime \prime}(\omega)$ is given by (\ref{LossModexemp3}). However, it is possible obtain an approximation of $G(t)$ using an appropriate numerical method to approximate the integral that defines $G(t)$. 

\medskip
    
The complex compliance is given by
    \begin{equation}
     \label{compilanceexamp3} J^{*}(w)=\frac{1}{G^*(w)}=\frac{1+\ln(iw)}{eiw-1}+(iw)^{-\beta},
    \end{equation}
    and from (\ref{StorageCOMP}) and (\ref{LossCOMP}) the storage and loss compliance's are
    \begin{eqnarray}
    \label{SCexemp3}J'(w)=
  w^{-\beta}\cos\left(\frac{\pi}{2}\beta\right)+\frac{g_1(w)}{2+2e^2w^2},\\
\label{LCexemp3}J''(w)=
  w^{-\beta}\sin\left(\frac{\pi}{2}\beta\right)+\frac{g_2(w)}{2+2e^2w^2},
    \end{eqnarray}
 respectively, where the functions $g_1$ and $g_2$ are defined by (\ref{g1exemplo3}) and (\ref{g2exemplo3}), respectively.
    
From (\ref{CreepComp}) and using (\ref{SCexemp3}) it follows that the creep compliance is given by 
 \begin{eqnarray}
 \nonumber J(t) &=&
 \frac{\sqrt{\pi }}{2e} G_{1,3}^{2,1}\left(\frac{t^2}{4 e^2}\bigg\rvert
\begin{array}{c}
 0 \\
 0,0,\frac{1}{2} \\
\end{array}
\right)+
\frac{2}{\pi} t^{\beta-1} \sin \left(\beta \frac{\pi }{2}\right) \cos \left(\beta \frac{\pi }{2}\right)+
\\
\nonumber &+&
\frac{1}{2} \left(e^{\frac{t-e}{e}}-e^{-\frac{t+e}{e}}\right)-\frac{\text{Chi}\left(\frac{t}{e}\right) \sinh \left(\frac{t}{e}\right)-\text{Shi}\left(\frac{t}{e}\right) \cosh \left(\frac{t}{e}\right)+\sinh \left(\frac{t}{e}\right)}{e}.
 \end{eqnarray}      
\end{itemize}

\begin{figure}[!ht]
\centering
\begin{tabular}{cc}  \includegraphics[scale=0.62]{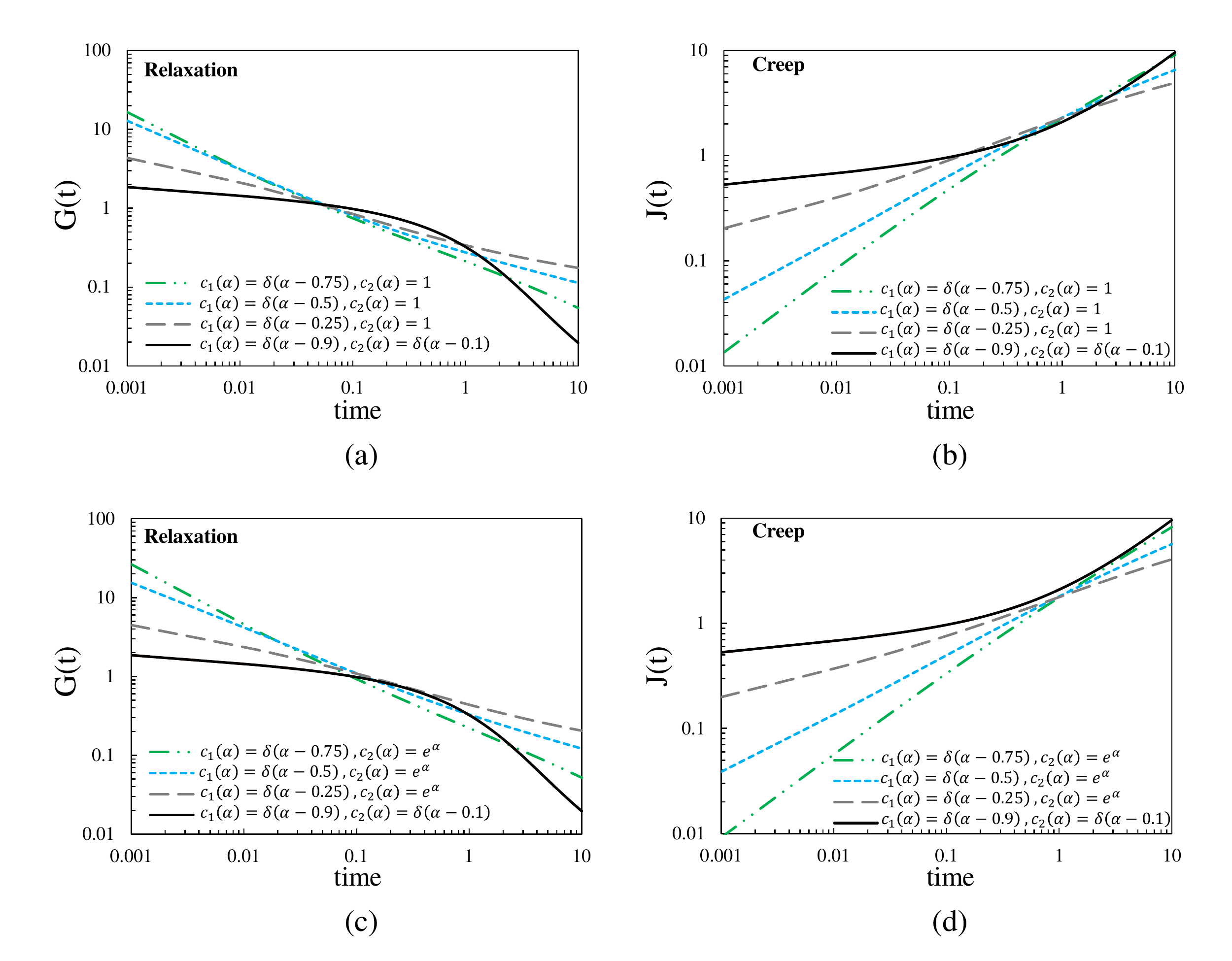}
\end{tabular}
\caption{Plots of the relaxation modulus and creep compliance (time is given in seconds) obtained for the Generalised Distributed-Order Maxwell Model (equation \ref{gamma2ele}). In (a) and (b) we consider $c_1(\alpha)=\delta(\alpha-\beta),~c_2(\alpha)=1$ (with $\beta=0.25;0.5;0.75$), and, $c_1(\alpha)=\delta(\alpha-0.9),~c_2(\alpha)=\delta(\alpha-0.1)$. In (c) and (d) we consider $c_1(\alpha)=\delta(\alpha-\beta),~c_2(\alpha)=e^{\alpha}$ (with $\beta=0.25;0.5;0.75$), and, $c_1(\alpha)=\delta(\alpha-0.9),~c_2(\alpha)=\delta(\alpha-0.1)$.}
\label{relCreep2}
\end{figure}

Figure \ref{relCreep2} shows the relaxation and creep functions for the different weighting functions considered before. For $c_1(\alpha)=\delta(\alpha-\beta),~c_2(\alpha)=1$ and $c_1(\alpha)=\delta(\alpha-\beta),~c_2(\alpha)=e^{\alpha}$ the relaxation modulus is obtained numerically.

The results are compared with the special case of the relaxation modulus and creep compliance (full line) obtained for the fractional Maxwell model - equation \ref{eqfracmaxwell} (see figure \ref{FMM}). It can be seen that the fractional Maxwell model appears to exhibit different behaviour compared to its distributed-order version. This is due to the fact that in the Generalised Distributed-Order Maxwell Model we have considered distributions that contain all the $\alpha$-order derivatives $\in [0,1]$ with similar weights. This way, the individual influence of each $\alpha$-order derivative is mitigated.

\medskip

It should be noted that it is difficult to determine a characteristic relaxation time of the model globally. The relaxation time depends strongly on the distribution $c(\alpha)$.

\section{Conclusions}

A generalised viscoelastic model for small deformations was derived from the Boltzmann theory. The model was further generalised by considering two distributed-order elements in series (as in the Maxwell model). The new model generalises the fractional viscoelastic model described in \cite{Schiessel1995}.

The new model presents a more general relaxation function, allowing its use for modelling a broader range of viscoelastic materials. Due the complexity of the mathematics involved, some of the relaxation functions must be obtained numerically.

The storage and loss modulus, the relaxation modulus and creep compliance were derived for special cases of $c(\alpha)$ (the function that distributes the weights through the different order derivatives);

The model can be easily generalised for large deformations (rheologically admissible constitutive equation) by using for example the Lodge rubber-like liquid form, combining the linear viscoelastic relaxation modulus presented in this work with finite strain kinematics \cite{Ng}. In the cases where an analytical expression is not possible to obtain for the relaxation modulus, a surrogate model or a meta-model can be used instead.

\smallskip

Due to the complexity of the mathematics involved and the difficulty in finding explicit expressions for general weighting functions, the research team is already working on the numerical implementation of this model in a general framework that will allow the study of the model behaviour in more complex geometries.

This will also allow to better understand the behaviour of the model for other weighting functions and to obtain a suitable mathematical expression for the characteristic relaxation time.

\section*{Acknowledgements}
This work was partially supported by the FCT-Funda\c{c}\~ao para a Ci\^encia e a Tecnologia (Portuguese Foundation for Science and Technology) through the project UIDB/00297/2020 (Centro de Matem{\'a}tica e Aplica\c{c}{\~o}es).

L.L. Ferr\'as would also like to thank FCT for financial support through CMAT (Centre of Mathematics of the University of Minho) projects UIDB/ 00013/2020 and UIDP/00013/2020.\\
\noindent
M.L. Morgado aknowledges funding by FCT through project UID/Multi/04621/2019 of CEMAT/IST-ID, Center for Computational and Stochastic Mathematics, Instituto Superior T{\'e}cnico, University of Lisbon.\\
\noindent


\newpage
\section*{Appendix A - Strain $(\gamma(t))$ as function of stress $(\sigma(t))$}

\subsection*{\textbf{Single-order case: $c(\alpha)=\delta(\alpha-\beta), 0<\beta<1$}}

In this case distributed order equation (\ref{distorder2}) reduces to 
 $ \sigma(t)=D^{\beta}\left( \gamma(t)\right)$, and therefore, we know, by applying the Riemann - Liouville integral, $I^{\beta} $, to both sides it follows, from \cite{Kilbas} 
\begin{equation}
    I^{\beta} \left(\sigma\right)(t)=I^{\beta}D^{\beta} \left(\gamma\right)(t)=\gamma(t)-\frac{1}{\Gamma(\beta)}\left(\frac{1}{\Gamma(1-\beta)}\int_0^t\frac{\gamma(s)}{(t-s)^{\beta}}ds\right)\bigg\rvert_{t=0}t^{\beta-1}.\end{equation}
The last term in the above equality vanishes if $\gamma(t)$ is bounded for $t\rightarrow 0^+$ and we obtain 
\begin{equation}
    \label{eq3}\gamma(t)=    I^{\beta} \left(\sigma\right)(t)=\frac{1}{\Gamma(\beta)} \int_{0}^{t}(t-s)^{\beta-1} \sigma(s) ds.
\end{equation}

\vspace*{0.5cm}    
On the other hand, using Theorem \ref{teorema2}, in this case, we have
$$
b(s)=\int_{0}^{1} \delta(\alpha-\beta) s^{\alpha} d \alpha=s^{\beta},
$$
therefore,
 $\ds b(re^{i\pi})= r^{-\beta}e^{i\pi\beta }=r^{-\beta}\left(\cos(\pi \beta)+i\sin(\pi \beta)\right)$ and by using Titchmarsh theorem, we have 
$$
\mathcal{L}^{-1}\left(\frac{1}{b(s)} ; t\right)=
\frac{1}{\pi}\int_0^{+\infty} e^{-rt} \sin(\pi\beta)r^{-\beta }dr=\frac{t^{\beta-1}}{\Gamma(\beta)}.
$$ 
Hence, by using (\ref{defgamma}) it follows 
\begin{equation}\label{expgammacase1}\gamma(t)=\sigma(t)\ast \frac{t^{\beta-1}}{\Gamma(\beta)}=
\frac{1}{\Gamma(\beta)}\int_0^t (t-s)^{\beta-1}\sigma(s)ds.\end{equation}
The result coincides with the Riemann-Liouville integral of order $\beta$ of $\sigma(t)$, as we saw before.

\subsection*{\textbf{Uniformly distributed-order case: $\displaystyle c(\alpha)=1$.}}
\ \noindent

In this case $$ b(s)=\int_{0}^{1} s^{\alpha} d\alpha=\frac{s-1}{\ln(s)}.$$

Let us determine the expressions for $\rho=\rho(r)$ and $\gamma=\gamma(r)$. \\ We have
\begin{eqnarray}
\label{ig1case2} b\left(r e^{i \pi}\right)&=&\frac{r e^{i \pi}-1}{\ln \left(r e^{i \pi}\right)}=\frac{-r-1}{\ln (-r)}=-\frac{(r+1)(\ln(r) -i \pi)}{(\ln(r))^2 + \pi^2} .
\end{eqnarray}
Using the fact that $$\ln(r)-i\pi=\sqrt{\log^2(r)+\pi^2}\, e^{-i\left(\frac{\pi}{2}+ \arctan(-\log(r)/\pi)\right)}$$ it follows 
\begin{eqnarray}
\nonumber b\left(r e^{i \pi}\right)&=&\frac{-(r+1) e^{-i\left(\frac{\pi}{2}+ \arctan(-\log(r)/\pi)\right) }}{\sqrt{\log^2(r)+\pi^2}}=\frac{(r+1) e^{i\left(\pi-\frac{\pi}{2}-\arctan(-\log(r)/\pi)\right) }}{\sqrt{\log^2(r)+\pi^2}}\\
\nonumber &=&\frac{(r+1)}{\sqrt{\log^2(r)+\pi^2}} 
\left(\cos\left( \frac{\pi}{2}-\arctan(-\log(r)/\pi)\right)
+i\sin\left( \frac{\pi}{2}-\arctan(-\log(r)/\pi)\right)
\right)\end{eqnarray}
Comparing with (\ref{aux}), we obtain
$$
\begin{array}{l}
\ds \rho\equiv \rho(r)=\frac{r+1}{\sqrt{ (\ln(r))^{2}+\pi^{2}}} \\
\ds \theta\equiv \theta(r)=\frac{1}{2}-\frac{1}{\pi} \operatorname{arctg}\left(-\frac{\ln(r)}{\pi}\right).
\end{array}
$$
Substituting in (\ref{aux1}) we obtain
\begin{eqnarray}\nonumber 
F(t)&=&\frac{1}{\pi} \int_{0}^{+\infty} e^{-r t} \frac{\sqrt{\operatorname{\ln(r)}^{2}+\pi^{2}}}{r+1} \operatorname{sin}\left(\frac{\pi}{2}-\operatorname{arctg}\left(-\frac{\ln(r)}{\pi}\right)\right) dr.\\
\nonumber &=&\int_{0}^{+\infty} \frac{e^{-r t}}{r+1} d r=e^{t} \Gamma(0, t), 
\end{eqnarray}
where $\ds \Gamma(0, t)$ is the incomplete Gamma function given by 
\begin{equation}
    \label{incompleteGamaFunc}\Gamma(0, t)=\int_{t}^{+\infty} t^{-1} e^{-t} d t
.
\end{equation}
From Theorem \ref{teorema2} we obtain the expression of $\gamma(t)$ 
\begin{equation}\label{gammacaso2}
    \gamma(t)=\int_{0}^{t} \sigma(s) e^{t-s} \Gamma(0, t-s) ds.
\end{equation}

\subsection*{\textbf{Exponential case: $\displaystyle c(\alpha)=e^{\alpha}$.}}
\ \noindent

In this case $$ b(s)=\int_{0}^{1} s^{\alpha} e^{\alpha} d\alpha=\frac{es-1}{1+\ln(s)}.$$

Let us determine the expressions for $\rho=\rho(r)$ and $\gamma=\gamma(r)$. \\ We have
\begin{eqnarray}
\label{ig1caseexp} b\left(r e^{i \pi}\right)&=&\frac{er e^{i \pi}-1}{1+\ln \left(r e^{i \pi}\right)}.
\end{eqnarray}
Since 
\begin{eqnarray}
\nonumber && er e^{i \pi}-1=(er+1)e^{i\pi},
\\
\nonumber && 1+\ln \left(r e^{i \pi}\right)= 1+\ln \left(r\right)+i \pi =\sqrt{(\ln(r)+1)^2+\pi^2}e^{i\left(\frac{\pi}{2}-\arctan(\frac{\ln(r)+1}{\pi})\right)},
\end{eqnarray}
 it follows 
\begin{eqnarray}
\nonumber b\left(r e^{i \pi}\right)&=&
\frac{(er+1) }{\sqrt{(\ln(r)+1)^2+\pi^2}}e^{i\left(\frac{\pi}{2}+ \arctan\left(\frac{\ln(r)+1}{\pi}\right)\right)}.
\end{eqnarray}
Comparing the above expression with (\ref{aux}), we obtain
$$
\begin{array}{l}
\ds \rho\equiv \rho(r)=\frac{(er+1) }{\sqrt{(\ln(r)+1)^2+\pi^2}}, \\
\ds \theta\equiv \theta(r)=\frac{1}{2}-\frac{1}{\pi} \operatorname{arctg}\left(-\frac{\ln(r)+1}{\pi}\right).
\end{array}
$$
Substituting in (\ref{aux1}) we obtain
\begin{eqnarray}\nonumber 
F(t)=\frac{e^{t/e}}{e}\Gamma(0,t/e),\,t>0.
\end{eqnarray}
where $\ds \Gamma(0, t)$ is the incomplete Gamma function given by (\ref{incompleteGamaFunc}).\\
From Theorem \ref{teorema2} we obtain the expression of $\gamma(t)$ 
\begin{equation}\label{gammacaso2}
    \gamma(t)=\int_{0}^{t} \sigma(s) e^{t-s} \Gamma(0, t-s) ds.
\end{equation}

\end{document}